# PERFORMANCE ANALYSIS FOR BANDWIDTH ALLOCATION IN IEEE 802.16 BROADBAND WIRELESS NETWORKS USING BMAP QUEUEING

Said EL KAFHALI   Abdelali EL BOUCHTI   Mohamed HANINI   and   Abdelkrim HAQIQ

Computer, Networks, Mobility and Modeling laboratory
e-NGN research group, Africa and Middle East
FST, Hassan 1st University, Settat, Morocco
{kafhalisaid, a.elbouchti, haninimohamed, ahaqiq}@gmail.com

***ABSTRACT***

*This paper presents a performance analysis for the bandwidth allocation in IEEE 802.16 broadband wireless access (BWA) networks considering the packet-level quality-of-service (QoS) constraints. Adaptive Modulation and Coding (AMC) rate based on IEEE 802.16 standard is used to adjust the transmission rate adaptively in each frame time according to channel quality in order to obtain multi-user diversity gain. To model the arrival process and the traffic source we use the Batch Markov Arrival Process (BMAP), which enables more realistic and more accurate traffic modelling. We determine analytically different performance parameters, such as average queue length, packet dropping probability, queue throughput and average packet delay. Finally, the analytical results are validated numerically.*

## 1. INTRODUCTION

### 1.1. Reference system

IEEE 802.16 standard networks accommodate the increasing user demand to enable pervasive, high-speed mobile internet access to a very large coverage area. Worldwide Interoperability for Microwave Access (WiMAX), first standardised in 2004 [1] known as IEEE 802.16, can provide broadband communications over wireless for various types of multimedia traffic, such as video streaming, VoIP, FTP etc.

WiMAX presents a very challenging multiuser communication problem [12] – many users in the same geographic area will require high on-demand data rates in a finite bandwidth, with low latency. Multiple access techniques allow different users to share the available bandwidth by allotting each user some fraction of the total system resources. Due to the diverse nature of anticipated WiMAX traffic, and the challenging aspects of the system deployment (mobility, neighboring cells, and high required bandwidth efficiency), the multiple access problems are quite complicated in WiMAX.

The IEEE 802.16 standard defines two types of operating mode for sharing the wireless medium: Point-to-Multipoint (PMP) and Mesh. The PMP mode adopts a cellular architecture, in this mode subscriber stations are scattered in the cellule around a central base station. There are two directions: Downlink (from BS to SS) and Uplink (from SS to BS). Transmissions from SSs are directed to and coordinated by the BS. On the other hand, in Mesh mode, the nodes are organized ad hoc and scheduling is distributed among them.

The WiMAX standard [1] defines the physical layer specifications and the Medium Access Control (MAC) signaling mechanisms. IEEE 802.16 uses two types of the modulation systems: OFDM (Orthogonal Frequency Division Multiple) and OFDMA (Orthogonal Frequency Division Multiple Access). OFDMA [12], extended OFDM, to accommodate many users in the same channel at the same time, and it has been adopted as the physical layer transmission technology for IEEE 802.16 based broadband wireless networks.

## 1.2. Related works

In order to promise the quality of real-time traffic and allow more transmission opportunity for other traffic types, an Adaptive Bandwidth Allocation model (ABA) for multiple traffic classes in IEEE 802.16 worldwide interoperability for microwave access networks was studied in [17]. The aim of work in [28] is to show how to exploit adaptive bandwidth allocation to increase system utilization (for the system administrator) with controlled QoS degradation (for the users). Instead of only focusing on bandwidth utilization or blocking/dropping probability, two new user-perceived QoS metrics, degradation ratio and upgrade/degrade frequency, are proposed. A Markov model is then provided to derive these QoS metrics. Using this model, authors evaluate the effects of adaptive bandwidth allocation on user-perceived QoS and show the existence of trade-offs between system performance and user-perceived QoS. Mathematical tools were used in [7, 8 and 9] to study performances parameters of both the connection-level and the packet-level for a model using two Connection Admission Control (CAC) schemes considered at a subscriber station in a single-cell IEEE 802.16 environment in which the base station allocates sub-channels to the subscriber stations in its coverage area.

For wireless mobile networks, the problem of providing packet-level QoS was studied quite extensively in the literature. A scheduling mechanism for downlink transmission was proposed in [32] to provide delay guarantee. In [15], authors proposed two credits based scheduling schemes which can efficiently serve real time burst traffic with reduced latency. The effect of the proposed schemes on latency, bandwidth utilization and throughput for real time burst flows is compared with Round Robin scheduling scheme. In [4], the proposed intergraded model can be applied to IEEE.16e. This model supports quality of service for request mechanism and data transmission in the uplink phase in the presence of channel noise; the authors calculate the performance parameters for single and multichannel wireless networks, like the requests throughput, data throughput and the requests acceptance probability and data acceptance probability. In [33], a dynamic fair resource allocation scheme was proposed to support real-time and non-real-time traffic in cellular CDMA networks. In [34], authors considered a data transmission system over a wireless channel, where packets are queued at the transmitter. Using A Markov approximation, they studied the statistics of the packet dropping process due to buffer overflow under automatic repeat request (ARQ) based error control scheme.

In [27], the authors consider a point-to-point wireless transmission where link layer ARQ is used to counteract channel impairments. They presented an analytical model framework to compute link-layer packet delivery delay statistics as a function of the packet error rate. An adaptive cross-layer scheduler was proposed in [14] for multiclass data services in wireless networks. The proposed scheduler uses the queuing information as well as it takes the physical layer parameters into account so that the required QoS performances can be achieved. The capacity of TDMA and CDMA-based broadband cellular wireless systems was derived in [30] under constrained packet-level QoS.

In [13], an analytical model is proposed to study the impacts of the channel access parameters, bandwidth configuration and piggyback policy on the performance. The impacts of physical burst profile and non-saturated traffic have also been taken into account. It is observed by simulations that the bandwidth utilization can be improved if the bandwidth for random channel access can be properly configured according to the channel access parameters, piggyback policy and network traffic. Besides, there isn't a single set of configurations that is always the best for all the network scenarios.

The authors in [5] present a pipeline approach to grant bandwidth at the BS of an IEEE 802.16 FDD network with half-duplex SSs. Based on this, they proposed a grant allocation algorithm, namely, the Half-Duplex Allocation (HDA) algorithm, which always produces a feasible grant allocation provided that the sufficient conditions are met. Although there have been several proposals for QoS scheduling frameworks and algorithms in IEEE 802.16 BWA networks in the literature [24, 31], they mainly focus on the QoS architecture and scheduling algorithm in a base station to satisfy diverse QoS requirements, rather than bandwidth request algorithm in a subscriber station.

A previous researcher in an attempt to address bandwidth allocation in IEEE 802.16 was reported by the authors in [10]. They considered a similar model in OFDMA based-WiMAX but they modeled packet-level by MMPP process and they compared various QoS measures.

Since the introduction of Batch Markovian Arrival Process (BMAP) by Lucantoni [11], the researchers [16, 21, and 29] prove that BMAP enables more realistic and more accurate traffic modeling; it can also capture dependency in traffic processes and outperforms MMPP and Poisson traffic models.

Since the incoming traffic in IEEE 802.16 has a self-similarity and a bursting nature causing correlation in inter-arrival times -which influences the performance of the system- we are motivated for using BMAP which can model such traffic correlation.

### 1.3. Aims of the paper

In this paper, we present a performance analysis for bandwidth allocation in IEEE 802.16 broadband wireless access networks considering the packet-level quality-of-service (QoS). Adaptive modulation and coding (AMC) rate based on IEEE 802.16 standard is used to adjust the transmission rate adaptively in each frame time according to channel quality in order to obtain multi-user diversity gain. A queueing analytical model is developed based on a Discrete-Time Markov Chain (DTMC) which captures the system dynamics in terms of the number of packets in the queue. We assume that the arrival process is modelled by the Batch Markov Arrival Process (BMAP) as the traffic source. Based on this model, various performance parameters such as average queue length, packet dropping probability due to lack of buffer space, the queue throughput, and the average queueing delay are obtained. Finally, the analytical results are validated by numerical results.

### 1.4. Organisation of the paper

The rest of the paper is organized as follows: In Section 2, we briefly introduce QoS architecture of IEEE 802.16 networks. Section 3 presents Modulation and Coding Schemes for IEEE 802.16. Section 4 describes the system model. The formulation of the analytical model is presented in Section 5. In section 6, different performance parameters are analytically determined. Section 7 states numerical results. Finally, section 8 gives a conclusion of this paper.

## 2. QOS ARCHITECTURE OF IEEE 802.16 NETWORKS

In this paper, we consider a point-to-point wireless mode (PMP) of IEEE 802.16, where a base station (BS) serves a set of subscriber stations (SSs). The Uplink and the downlink are served in the separate region of physical layer (OFDMA/TDD) frame .the downlink channel is in broadcast mode, but an SS is only required to process data which are addressed to itself. In the uplink sub-frame, on the other hand, the SSs transmit data to the BS in a Time Division Multiple Access (TDMA) manner. Downlink and uplink sub-frames are duplexed using one of the following techniques: (Frequency Division Duplex FDD), where downlink and uplink sub-frames occur simultaneously on separate frequencies, and Time Division Duplex (TDD), where downlink and uplink sub-frames occur at different times and usually share the same frequency. SSs can be either full duplex or half-duplex.

IEEE 802.16e uses a connection-oriented medium access control (MAC) protocol which provides a mechanism for the SSs to request bandwidth to the BS. IEEE 802.16e MAC supports two classes of SS: grant per connection (GPC) and grant per SS (GPSS). In the case of GPC, bandwidth is granted to a connection individually. In contrast, for GPSS, a portion of the available bandwidth is granted to each of the SSs and each SS is responsible for allocating bandwidth among the corresponding connections.

The lengths of the downlink and uplink sub-frames for each SS are determined by the BS and broadcast to the SSs through downlink and uplink map messages (UL-MAP and DL-MAP) at the beginning of each frame. Therefore, each SS knows when and how long to receive from and transmit data to the BS. In the uplink direction, each SS can request bandwidth to the BS by using BW-request packets.

WiMAX is associated with the IEEE 802.16 standard [1, 2, and 3], which defines five classes of traffic flows representing different types of services in the following order: Unsolicited Grant Service (UGS), Extended Real Time Polling Service (ertPS), Real Time Polling Service (rtPS), Non-Real Time Polling Service (nrtPS), and Best Effort Service (BE). Each class has its QoS mechanisms at the Media Access Control (MAC) layer to support the various applications. UGS is designed to support real-time service flows that generate fixed-size data packets on a periodic basis, such as VoIP without silence suppression. ertPS supports real-time applications which generate variable-sized data packets periodically that require guaranteed data rate and delay with silence suppression. rtPS supports real-time service flows that generate variable data packets size on a periodic basis. nrtPS supports delay-tolerant data streams which are more bursty in nature, such as FTP, in general, the nrtPS can tolerate longer delays and is insensitive to delay jitter, but requires a minimum throughput. BE supports traffic with no QoS requirements, such as email, and therefore may be handled on a resource-available basis.

## 3. Modulation and Coding Schemes for IEEE 802.16

Adaptive modulation and coding scheme (AMC) is supported in the WiMAX networks. The basic idea of AMC is to maximize the data rates by adjusting the transmission parameters according to the fluctuations in the channel.

The channel quality is determined by the instantaneous received Signal-to-Noise Ratio (SNR) $\gamma$ in each time slot. We assume that the channel is stationary over the transmission frame time. Lower data rates are achieved by using Modulation Level and rate error correcting corresponding to Rate $ID = 0$ (BPSK and 1/2). The higher data rates are achieved by using Modulation Level and rate error correcting corresponding to Rate $ID = 6$ (64QAM and 3/4). In all, there are 52 different possible configurations of modulation order and coding types and rates [12], although most implementations of WiMAX will offer only a fraction of these. Table 1 lists these schemes represented by different rate IDs for IEEE 802.16 WiMAX Networks.

In an OFDMA system [12], each user will be allocated a block of subcarriers, each of which will have a different set of SNR. Therefore, care needs to be paid to which constellation/coding set is chosen based on the varying SNR across the subcarriers.

To determine the mode of transmission (i.e., modulation level and coding rate), an estimated value of SNR at the receiver is used. In this case, the SNR at the receiver is divided into $N+1$ nonoverlapping intervals (i.e., $N=7$ in WiMAX) by thresholds $\Gamma_n (n \in \{0,1,...,N\})$ where $\Gamma_0 < \Gamma_1 < ... \Gamma_{N+1} = \infty$. The subchannel is said to be in state n (i.e., *rate* $ID = n$ will be used) if $\Gamma_n \le \gamma < \Gamma_{n+1}$. To avoid possible transmission error, no packet is transmitted when $\gamma < \Gamma_0$. Note that, these thresholds correspond to the required SNR specified in the WiMAX standard, SNR, Signal-to-Noise Ratio [18, 23, and 25].

That is, $\Gamma_0 = 6.4,\ \Gamma_1 = 9.4, ...,\ \Gamma_N = 24.4$ (as shown in Table 1).

Table 1: IEEE 802.16 Profiles.

| *Rate ID* | *Modulation Level (Coding)* | *Information Bits/Symbol* | *Required SNR (db)* |
|---|---|---|---|
| 0 | BPSK (1/2) | 0.5 | 6.4 |
| 1 | QPSK (1/2) | 1 | 9.4 |
| 2 | QPSK (3/4) | 1.5 | 11.2 |
| 3 | 16QAM (1/2) | 2 | 16.4 |
| 4 | 16QAM (3/4) | 3 | 18.2 |
| 5 | 64QAM (2/3) | 4 | 22.7 |
| 6 | 64QAM (3/4) | 4.5 | 24.4 |

# 4. MODEL DESCRIPTION

## 4.1. Arrival Process Traffic

The BMAP has received considerable interest during the last few years. It was first introduced by Neuts [11] as the versatile Markovian point Process. It generalizes Markovian Arrival Process (MAP) introduced by Lucantoni et al. [20].

To capture the arrival process traffic, we use a BMAP process [6]. The arrivals in the BMAP is directed by the irreducible continuous time Markov chain CTMC with a finite state space {0, 1, …, *S*}. Sojourn time of the CTMC in the state *s* has exponential distribution with parameter $\lambda_s$. After time expires, with probability $p_0(s,s')$ the chain jumps into the state *s'* without generation of packets and with probability $p_k(s,s')$ the chain jumps into the state *s'* and a batch consisting of k packets is generated, $k \ge 1$. The introduced probabilities satisfy conditions: $p_0(s,s)=0,$ the sum of the probabilities of all outgoing transitions has to be equal to 1,

$$\sum_{k=1}^{\infty}\sum_{s'=0}^{S} p_k(s,s') + \sum_{\substack{s'=0 \\ s'\neq s}}^{S} p_k(s,s') = 1,\ 0 \le s \le S. \tag{1}$$

The BMAP is a two dimensional Markov process $\{A(t), J(t)\}$ on the state space $\{(i,j) / i \ge 0, 0 \le j \le S\}$ with infinitesimal generator given by:

$$\Psi = \begin{pmatrix} D_0 & D_1 & D_2 & D_3 & \cdots \\ 0 & D_0 & D_1 & D_2 & \cdots \\ 0 & 0 & D_0 & D_1 & \cdots \\ 0 & 0 & 0 & D_0 & \cdots \\ \vdots & \vdots & \vdots & \vdots & \ddots \end{pmatrix} \tag{2}$$

where the matrices $D_0 = [D_{ss'}],\ 0 \le s \le S, 0 \le s' \le S$ has negative diagonal elements and non negative off diagonal elements given by:

$$D_{ss'} = \begin{cases} -\lambda_s, & s' = s \\ \lambda_s p(s,s'), & s \neq s' \end{cases} \quad (3)$$

The matrices $D_k, k > 0$ are defined by:

$$\mathrm{D}_k = [D_{k,ss'}],\ 0 \leq s \leq \mathrm{S},\ 0 \leq s' \leq \mathrm{S},\ k > 0 \quad (4)$$

where:

$$D_{k,ss'} = \lambda_s p_k(s,s'),\ \ 0 \leq s \leq \mathrm{S},\ 0 \leq s' \leq \mathrm{S},\ k > 0. \quad (5)$$

The matrix $D$ defined by. $\mathrm{D} = \sum_{k=0}^{\infty} \mathrm{D}_k$ , is an irreducible infinitesimal generator. We also assume that $\mathrm{D}_k \neq D_0$, which ensures that arrivals will occur.

The variable $A(t)$ counts the number of arrivals during $[0,t[$ and the variable $J(t)$ represents the phase of the arrivals process.

The steady-state probability vector $\pi_{BMAP}$ of the CTMC with generator $D$ can be calculated as usual:

$$\pi_{BMAP}.\mathrm{D} = \vec{0}, \quad \pi_{BMAP}.e = 1. \quad (6)$$

Where $\vec{0}$ and $e$ are row and column vectors consisting of zeros and units, respectively.

The mean steady-state arrival rate generated by the BMAP is:

$$\lambda_{BMAP} = \pi_{BMAP} \sum_{k=1}^{\infty} k\mathrm{D}_k e. \quad (7)$$

More detail and results concerning this Process can be found in [19] for instance.

The probability $f_a(\lambda_s, T)$ of $a = 0,1,...,A$ incoming packets, with $A$ denoting the maximum packets' number, that arrive with mean rate $\lambda_s$ within a time slot interval $T$ is given by:

$$f_a(\lambda_s, T) = \frac{e^{-\lambda_s T}(\lambda_s T)^a}{a!} \quad (8)$$

It is also essential for the condition $\sum_{a=A}^{+\infty} \frac{e^{-\lambda_s T}(\lambda_s T)^a}{a!} < er\ \forall s$ always to stand, with $er$ expressing a sufficiently small number.

Note that the probability that $a$ Poison arrivals with average rate $\lambda_s$ occur during an interval T is given by the $S \times S$ diagonal matrix $\xi_a$ which is defined as:

$$\xi_a = \begin{bmatrix} f_a(\lambda_1, T) & & & \\ & f_a(\lambda_2, T) & & \\ & & \ddots & \\ & & & f_a(\lambda_s, T) \end{bmatrix} \quad (9)$$

## 4.2. System Model

We consider an infrastructure-based wireless access network, where connections are established between a base station (BS) and multiple subscribers stations (SSs) through a TDMA/TDD access mode using single carrier air-interface (as shown in Figure 1). Each subscriber station serves multiple connections. For each connection a separate queue in SS with size $X$ packets is used for buffering the packets from higher layers. In particular, for one connection, there is a queue for uplink and another queue for downlink transmissions from the SS and the BS, respectively. We consider an SS of type GPC. Therefore, through the SS a certain amount of bandwidth is reserved for each connection during bandwidth allocation.

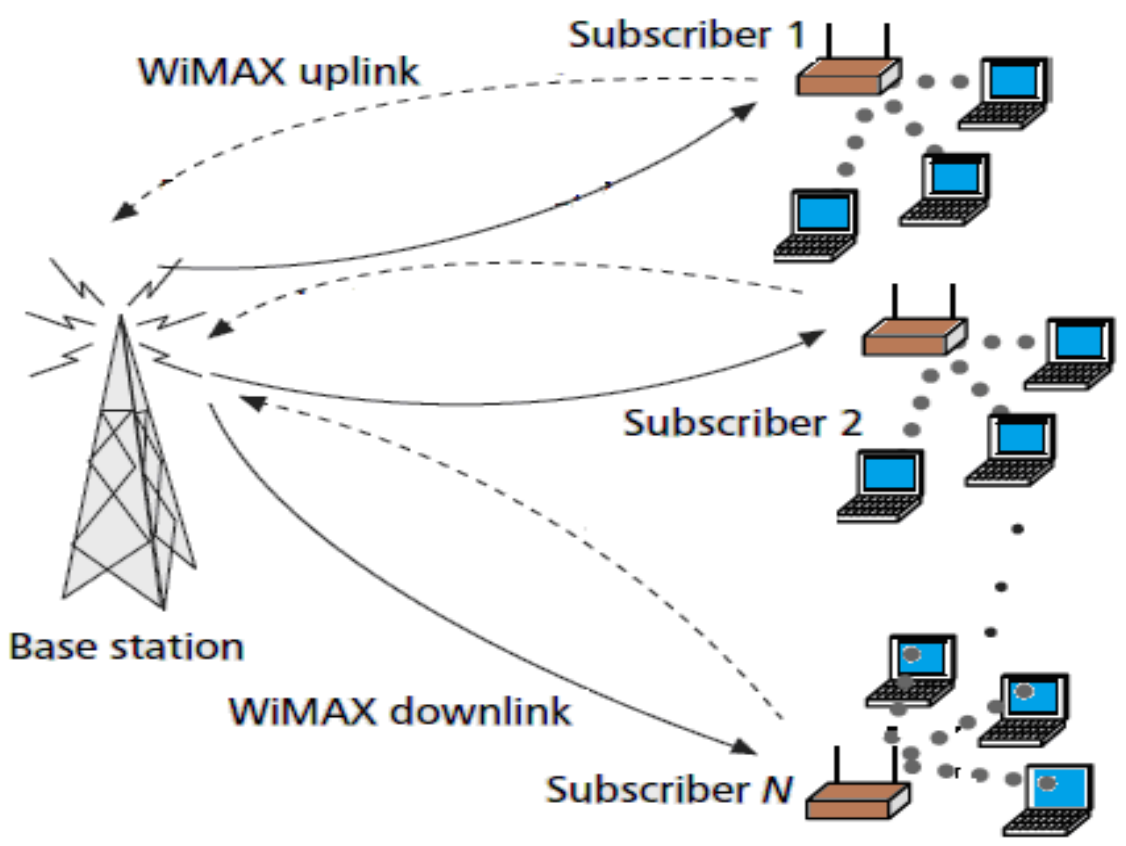

Figure 1: System model

# 5. Formulation of the Analytical Model

## 5.1. State Space

The state of the queue is observed at the beginning of each frame. We assume that connection $i$ is allocated with $b_i$ units of bandwidth and a packets arriving during frame period $f$ will not be transmitted until frame period $f+1$ at the earliest. The state space of the queue can be defined as follows:

$$E = \{(x, s); 0 \le x \le X, 1 \le s \le S\}, \tag{10}$$

where $x$, $s$ represent, the number of packets in the queue, the state (phase) of an irreducible continuous time Markov chain of BMAP arrival process respectively.

### 5.2. Transition Matrix for the Queue

The transition matrix $\mathrm{M}$ for the queue can be expressed as follows:

$$\mathrm{M} = \begin{bmatrix} m_{0,0} & m_{0,1} \cdots & m_{0,A} & & & & \\ \vdots & \vdots \ddots & \ddots & & & & \\ m_{D,0} & m_{D,1} \cdots & m_{D,D} & \cdots & \cdots & m_{D,D+A} & \\ \ddots & \ddots & \ddots & \ddots & \ddots & \ddots & \\ & m_{X-A,X-A-D} & \cdots & \cdots & & m_{X-A,X} & \\ & \ddots & \ddots & \ddots & \ddots & \vdots & \\ & & m_{X,X-D} & \cdots & \cdots & m_{X,X} & \end{bmatrix} \tag{11}$$

The rows of matrix $\mathrm{M}$ represent the number of packets in the queue and element $m_{x,x'}$ inside this matrix denotes the transition probability for the case when the number of packets in the queue changes from $x$ in the current frame to $x'$ in the next frame. Also, the maximum number of packets that can enter into or depart from the queue within a frame time is represented with $A$ and $D$ respectively.

The maximum number of transmitted packets within a time slot is given by $D^{'} = \min(x, D)$. Hence, if $k$ represents the number of the successfully transmitted packets. The probability that $k$ packets will be transmitted in a timeslot is obtained by the matrix $T_k$ which is defined as follows:

$$T_k = \begin{cases} \binom{x}{k}(1-\theta)^k \theta^{x-k} & , k < D^{'} \\ \sum_{j=U^{'}}^{x} \binom{x}{j}(1-\theta)^j \theta^{x-j} & , k = D^{'} \end{cases} \tag{12}$$

where $\theta$ is the probability that a packets is successfully transmitted.

The elements in the matrix $\mathrm{M}$ can be obtained as follows:

$$m_{x,x-u} = \Psi \times \sum_{k-a=u} \xi_a \times T_k \tag{13}$$

$$m_{x,x+v} = \Psi \times \sum_{a-k=v} \xi_a \times T_k \tag{14}$$

$$m_{x,x} = \Psi \times \sum_{k=a} \xi_a \times T_k \tag{15}$$

for $u = 1,2,...,D^{'}$ and $v = 1,2,...,A$ where, $k \in \{0,1,...,D^{'}\}$ and $a \in \{0,1,...,A\}$ represent the number of departed packets and the number of packets arrivals, respectively.

With $m_{x,x-u}$, $m_{x,x+v}$ and $m_{x,x}$ we represent the probability that the number of packets in the queue increases by $u$, decreases by $v$, and does not change, respectively.

The remaining rows $\{x = X-A+1, X-A+2,..., X\}$ of the matrix $\mathrm{M}$, include the occurrence where some packets would be dropped due to lack of queue space. We calculate the probabilities as follows:

$$m_{x,x+v} = \sum_{a=v}^{A} m^{'}_{x,x+a} \quad \text{for } x+v \geq X \tag{16}$$

Which express that no packet has been dropped and occur when more incoming packets to an already fully queue are dropped. Additionally, the last element of the main diagonal of $\mathrm{M}$ is given by:

$$m_{x,x} = m^{'}_{x,x} + \sum_{a=1}^{A} m^{'}_{x,x+a} \quad \text{for } x = X \tag{17}$$

where $m^{'}_{x,x}$ is obtained for the case without any packets dropping.

Equations (16) and (17) indicate the case that the queue will be full if the number of incoming packets is greater than the available space in the queue. In other words, the transition probability to the state that the queue is full can be calculated as the sum of all the probabilities that make the number of packets in queue equal to or larger than the queue size $X$.

## 6. PERFORMANCE PARAMETERS

The performance parameters are analytically calculated using the steady state probability of the system. The vector $\pi$ of these probabilities is obtained by solving the system $\pi.\mathrm{M} = \pi$ and $\pi.1 = 1$, where $1$ is a column matrix of ones.

The steady-state probabilities, denoted by $\pi(x,s)$ for the state that there are $x \in \{0,1,...,X\}$ packets in the queue, can be extracted from matrix $\pi$ as follows:

$$\pi(x,s) = [\pi]_{X \times S} \quad , \mathrm{s} = 1,...,S \tag{18}$$

The matrix $\pi$ contains the steady state probabilities corresponding to the number of packets in the queue and the state (phase) of an irreducible continuous time Markov chain of BMAP arrival process. Using the steady state probabilities, the various performance measures can be obtained.

### 6.1. Average Queue Length

The average number of packets in the transmission queue is obtained as follows:

$$\overline{X} = E(x) = \sum_{x=0}^{X} x \sum_{s=1}^{S} \pi(x,s) \tag{19}$$

### 6.2. Packet Dropping Probability

In order to compute the packet dropping probability ( $p_{drop}$ ), we firstly obtain the number of dropped packets per time slot, obtained using the average number of dropped packets per frame. If $x$ is the number of packets in the queue and this number increases by $n$, the number of dropped packets $X_{drop}$ is: $X_{drop} = n - (X - x).\mathbb{1}_{]X-n,+\infty[}(x)$ where

$$\mathbb{1}_A(x) = \begin{cases} 1 & if \ x \in A \\ 0 & otherwise \end{cases} \tag{20}$$

The average number of dropped packets per frame is obtained as follows:

$$\begin{aligned} \overline{X}_{drop} &= E(X_{drop}) \\ &= \sum_{x=0}^{X} \sum_{s=1}^{S} \sum_{n=X-x+1}^{A} \left( \sum_{l=1}^{S} [m_{x,x+n}]_{s,l} \right).(n-(X-x)).\pi(x,s) \end{aligned} \tag{21}$$

where the term $\left( \sum_{l=1}^{S} [m_{x,x+n}]_{s,l} \right)$ in Equation (21) indicates the total probability that the number of packets in the queue increases by $n$ at every arrival phase. The probability $m_{x,x+n}$ is used rather than the probability of packet arrival, because the packet transmission in the same frame is considered.

After calculating the average number of dropped packets per frame, we can obtain the probability that an incoming packet is dropped as follows:

$$p_{drop} = \frac{\overline{X}_{drop}}{\lambda_{BMAP}} \tag{22}$$

where $\lambda_{BMAP}$ is the mean steady state arrival rate generated by the BMAP (as obtained from (7)).

### 6.3. Queue throughput

It measures the number of packets transmitted in one frame and can be obtained from:

$$\varphi = \lambda_{BMAP}(1 - p_{drop}) \tag{23}$$

### 6.4. Average Packet Delay

The average packet delay is defined as the number of frames that a packet waits in the queue since its arrival before it is transmitted.
We use Little's formula [22] to obtain average packets delay as follows:

$$\overline{D} = \frac{\overline{X}}{\varphi} = \frac{\overline{X}}{\lambda_{BMAP}(1 - P_{drop})} \tag{24}$$

where $\varphi$ is the throughput and $\overline{X}$ is the average queue length.

## 7. Numerical Results

In the next, performance parameters are numerically evaluated, using Matlab software.

### 7.1. Parameter Setting

We consider the system model depicted in section 4. Adaptive Modulation and Coding (AMC) is used in which the modulation level and the coding rate are increased if the channel quality permits. Table 1 lists these schemes represented by different rate IDs for IEEE 802.16.
The maximum number of packets that can be transmitted in one frame period is 150 packets per frame.
The queue size is assumed to be 150 packets (i.e. $X = 150$).
For simplicity of computation we assume the maximum size of the batch to be 2, and the matrices governing the state transitions of the BMAP are giving as follows:

$$D_0 = \begin{pmatrix} -2 & \frac{1}{2} \\ \frac{1}{8} & -1 \end{pmatrix}, D_1 = \begin{pmatrix} \frac{1}{2} & \frac{1}{4} \\ \frac{1}{4} & \frac{1}{4} \end{pmatrix}, D_2 = \begin{pmatrix} \frac{1}{4} & \frac{1}{2} \\ \frac{1}{4} & \frac{1}{8} \end{pmatrix}$$

Its sojourn times at each state are assumed to be exponentially distributed with rates $\lambda_0 = 2.0$, $\lambda_1 = 1.0$ respectively.
The performance parameters are measured respectively under different amount of allocated bandwidth $b$, under different channel qualities with constant traffic intensity, and under different traffic intensities with channel SNR in the range of rate $IDn = 0$.
In this work, bandwidth $b$ is defined (as in [10]) as the number of packets that can be transmitted in one frame using rate $IDn = 0$.

### 7.2. Results and Discussion

We first examine the impact of traffic intensity on bandwidth allocation. Variations in throughput with traffic intensity are shown in Figure 2. When the traffic intensity increases, the throughput increases until it becomes saturated. At this point (e.g., $3.0$), the arriving packets cannot be transmitted faster than the transmission rate that the channel quality allows.

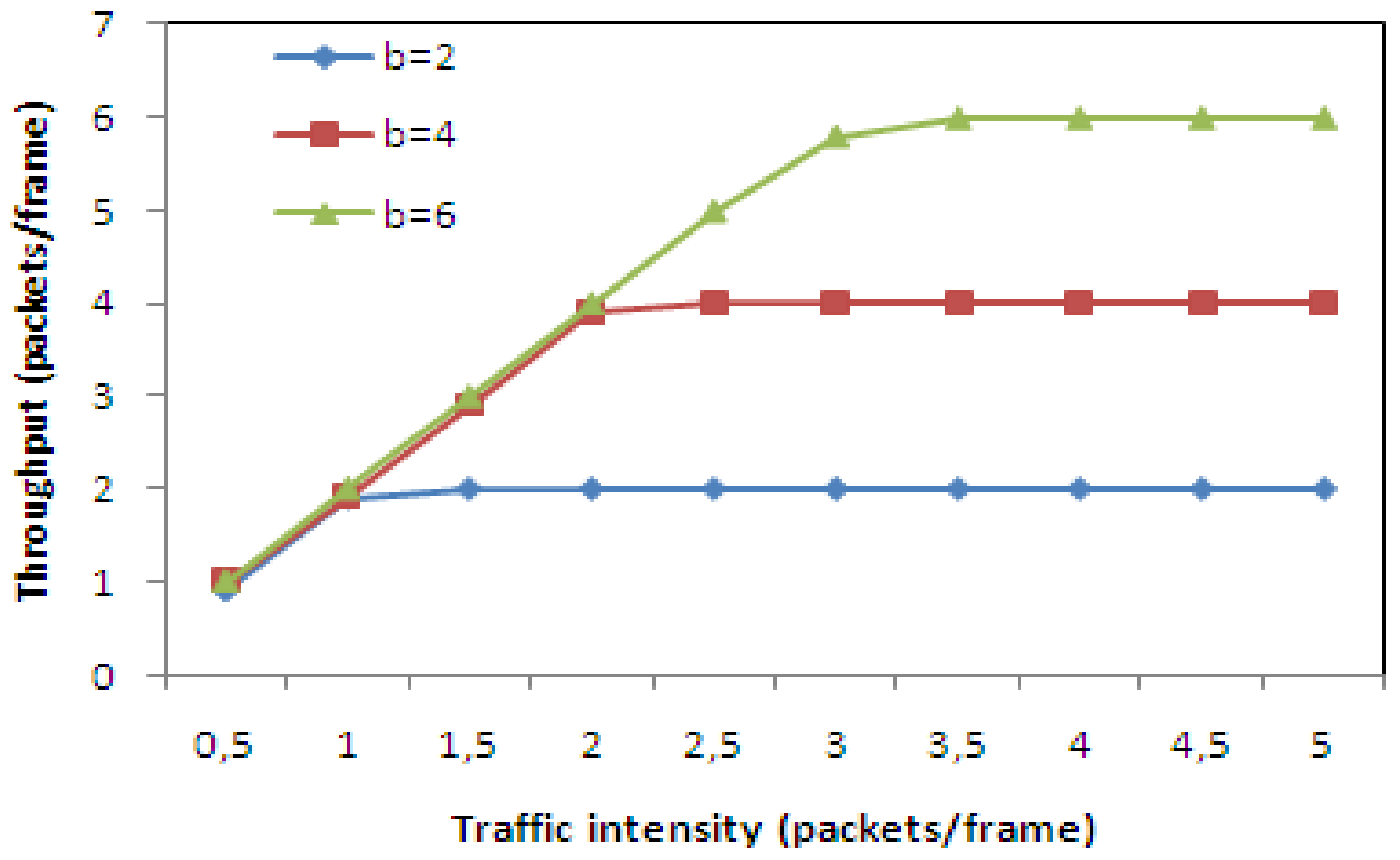

Figure 2: Throughput under traffic intensity.

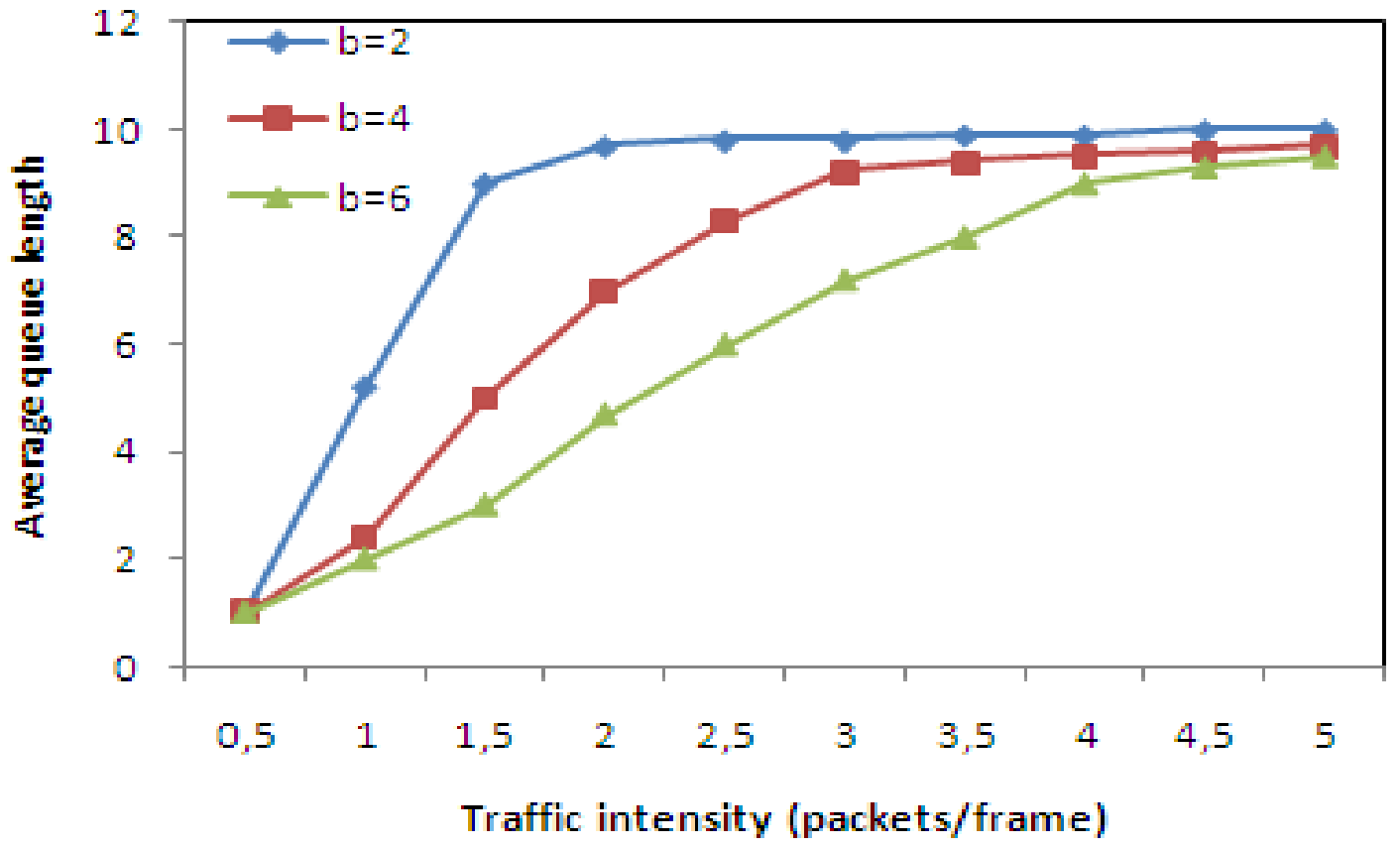

Figure 3: Average queue length under traffic intensity.

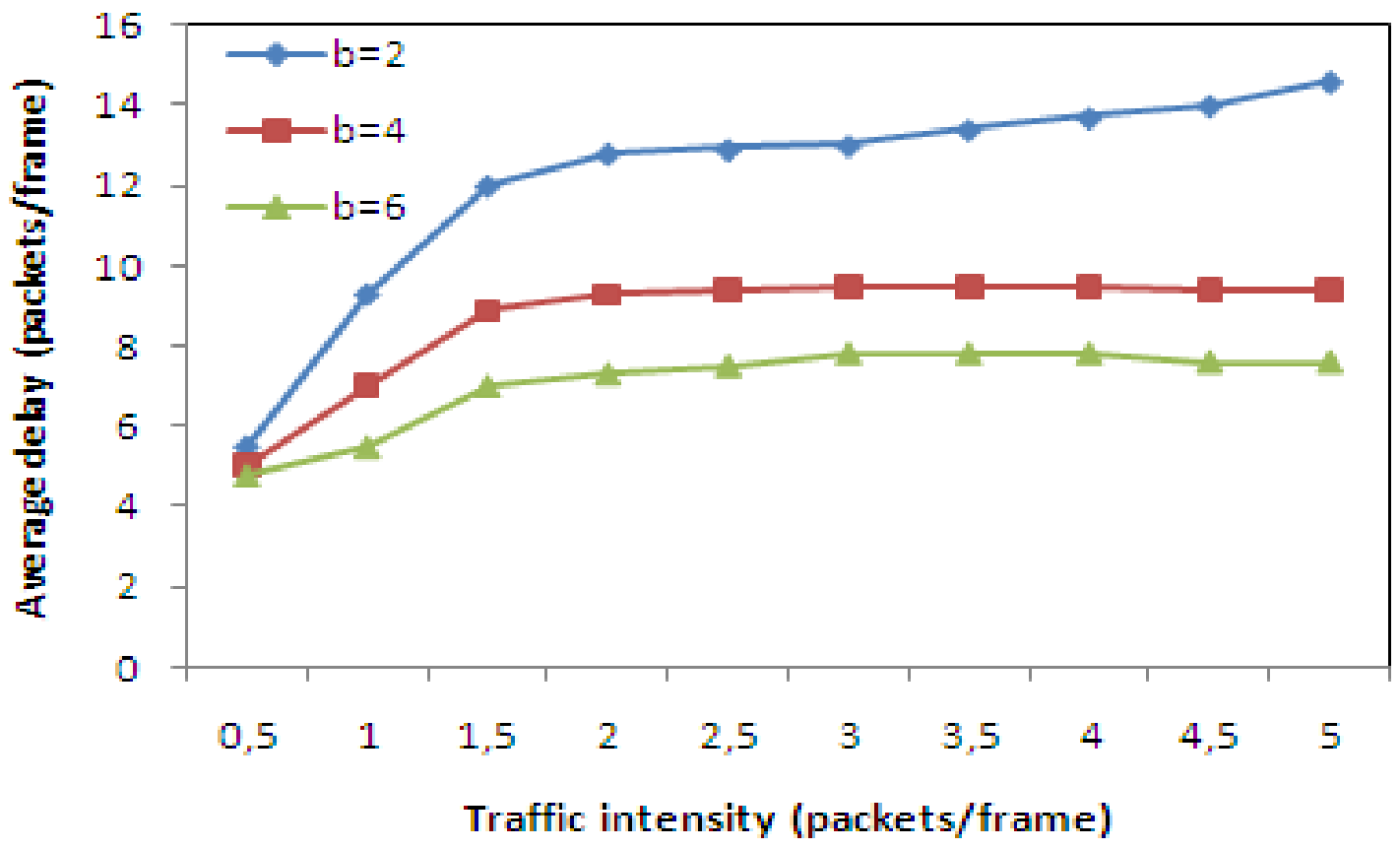

Figure 4: Average delay under traffic intensity.

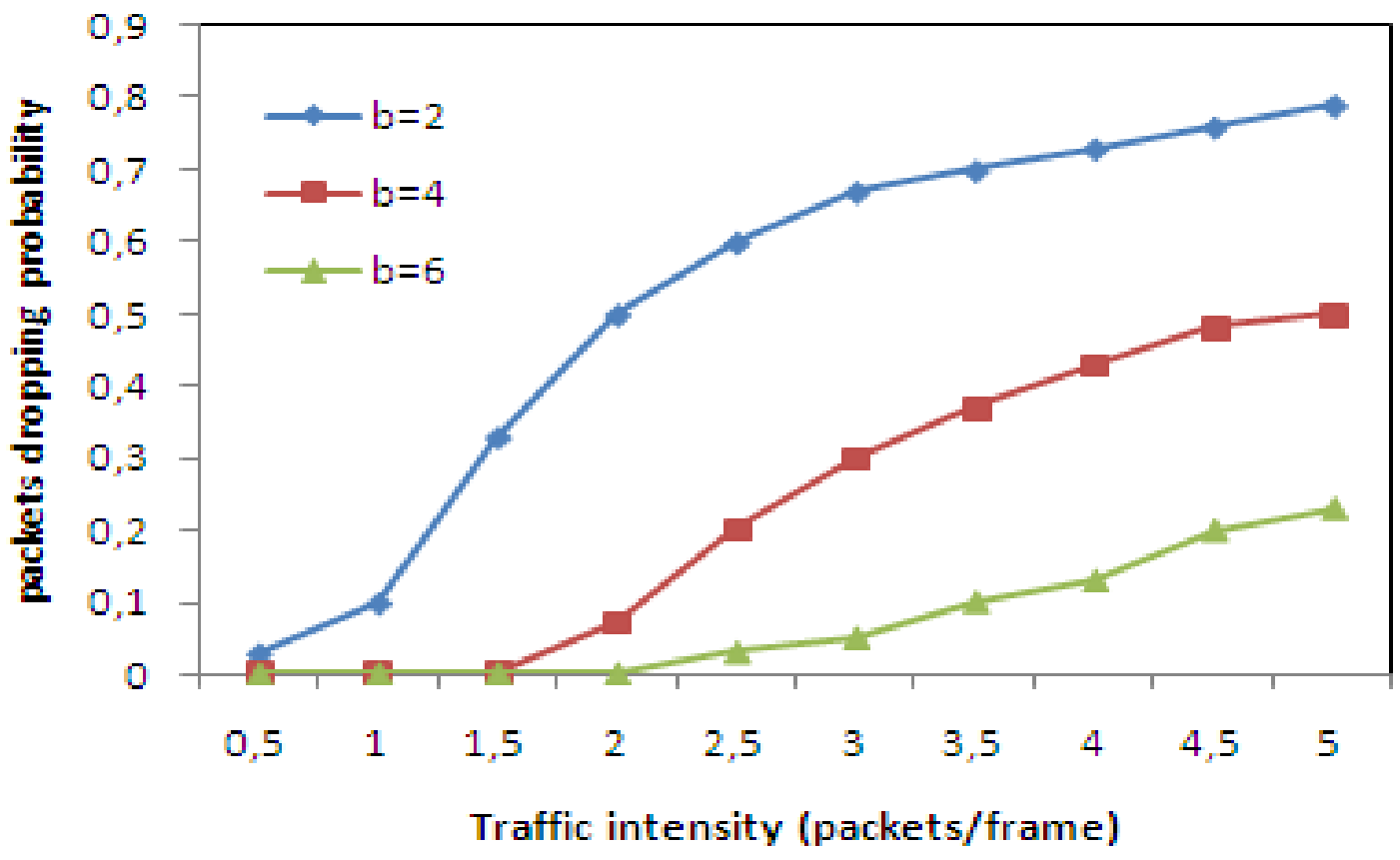

Figure 5: packets dropping probability under traffic intensity.

Average queue length, average delay and packets dropping probability increase as the traffic intensity increases (Figures 3, 4 and 5). Therefore, we can say that the increase of the parameters previously mentioned is linked with the traffic intensity. On the other hand, those parameters decrease as the channel quality improves (Figures 6, 7 and 8).

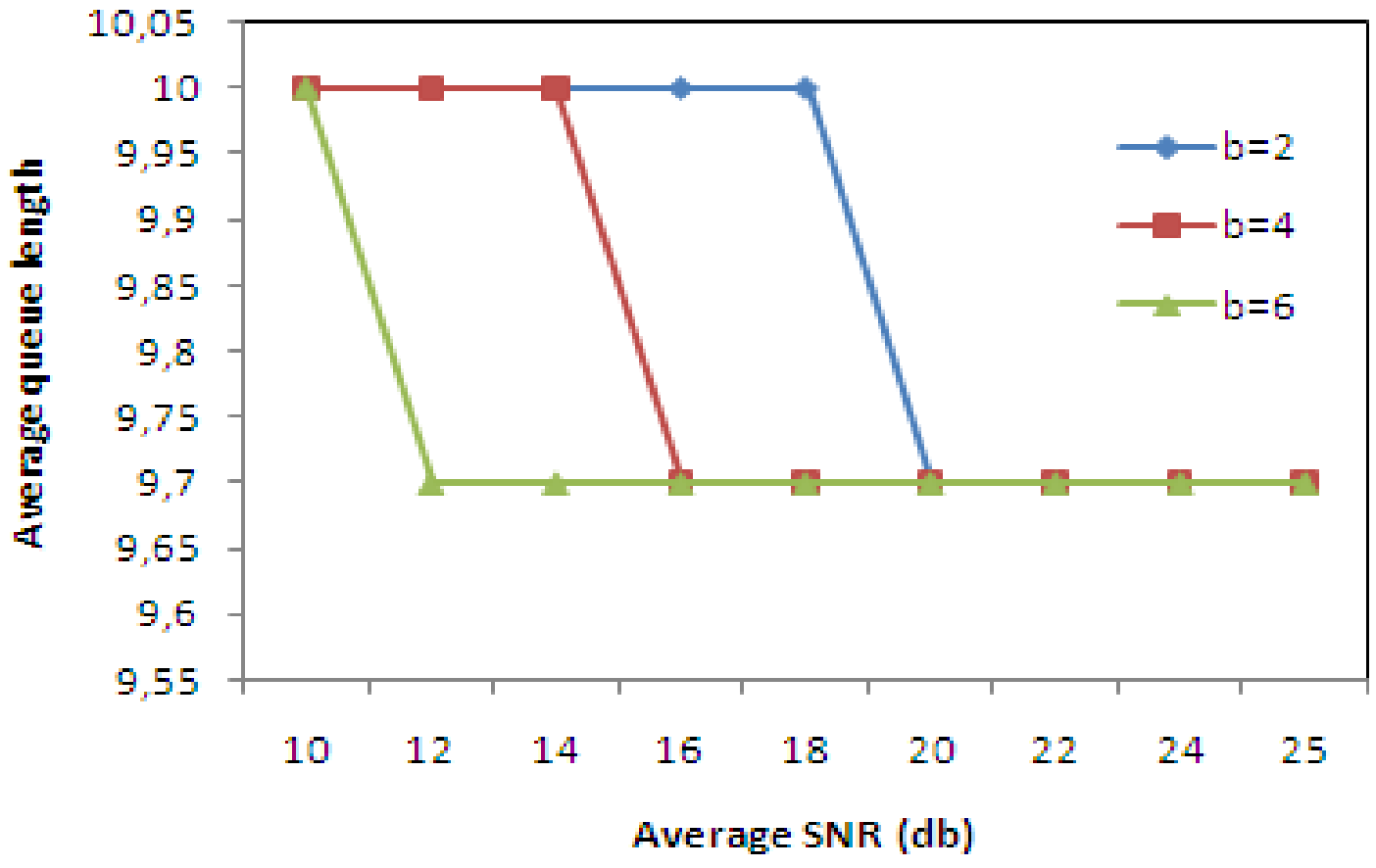

Figure 6: Average queue length under different channel qualities.

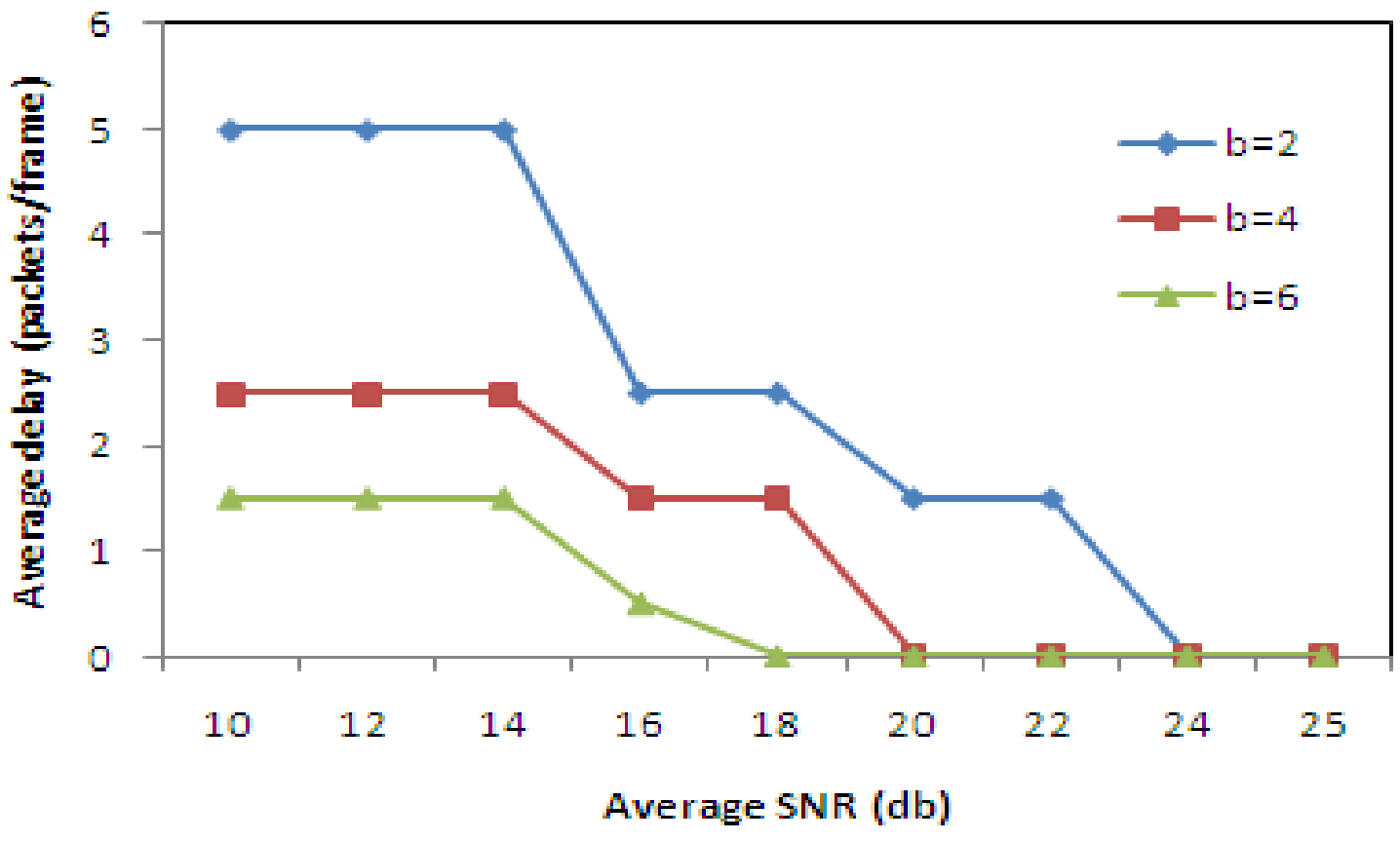

Figure 7: Average delay under different channel qualities.

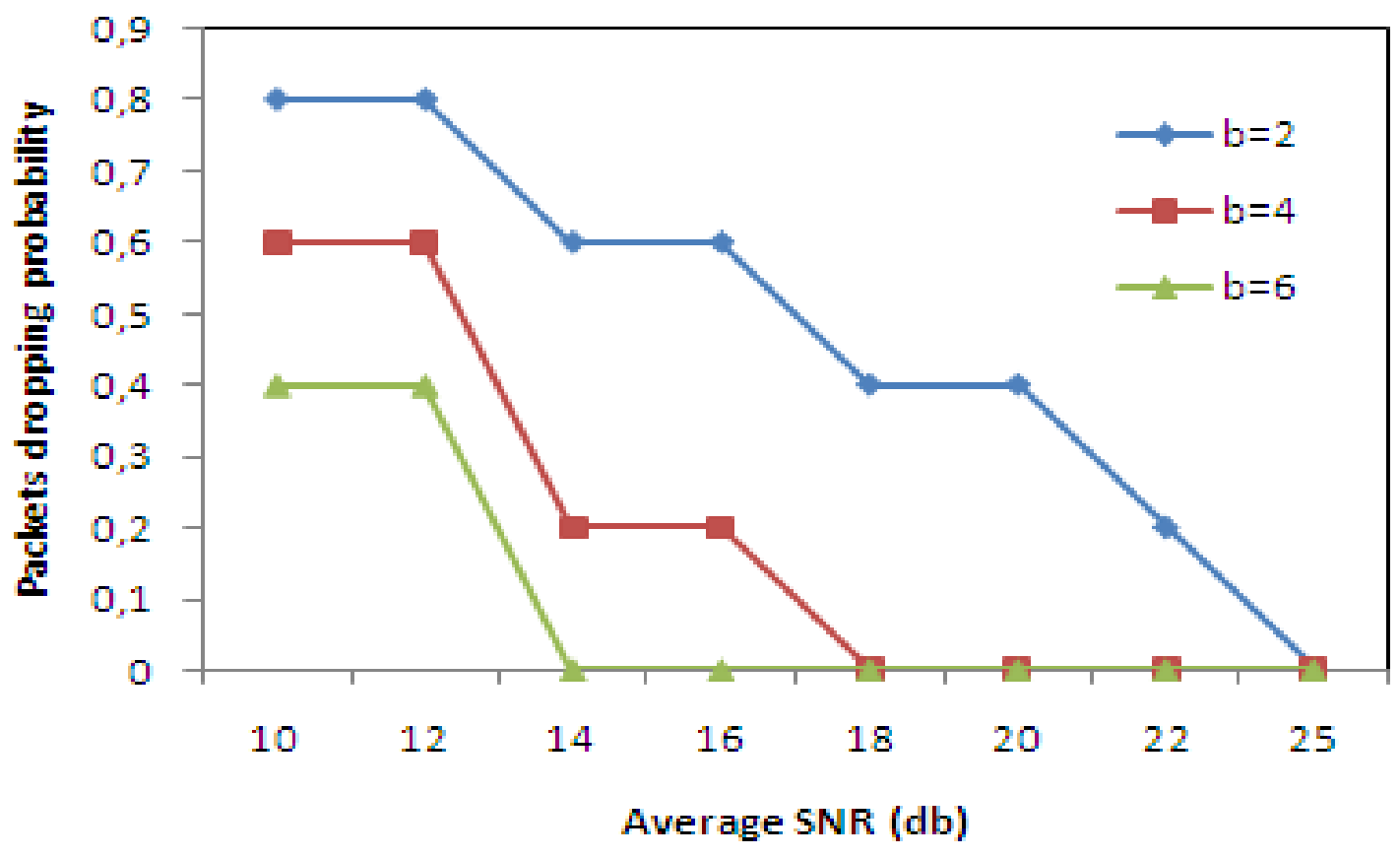

Figure 8: Packet dropping probability under different channel qualities.

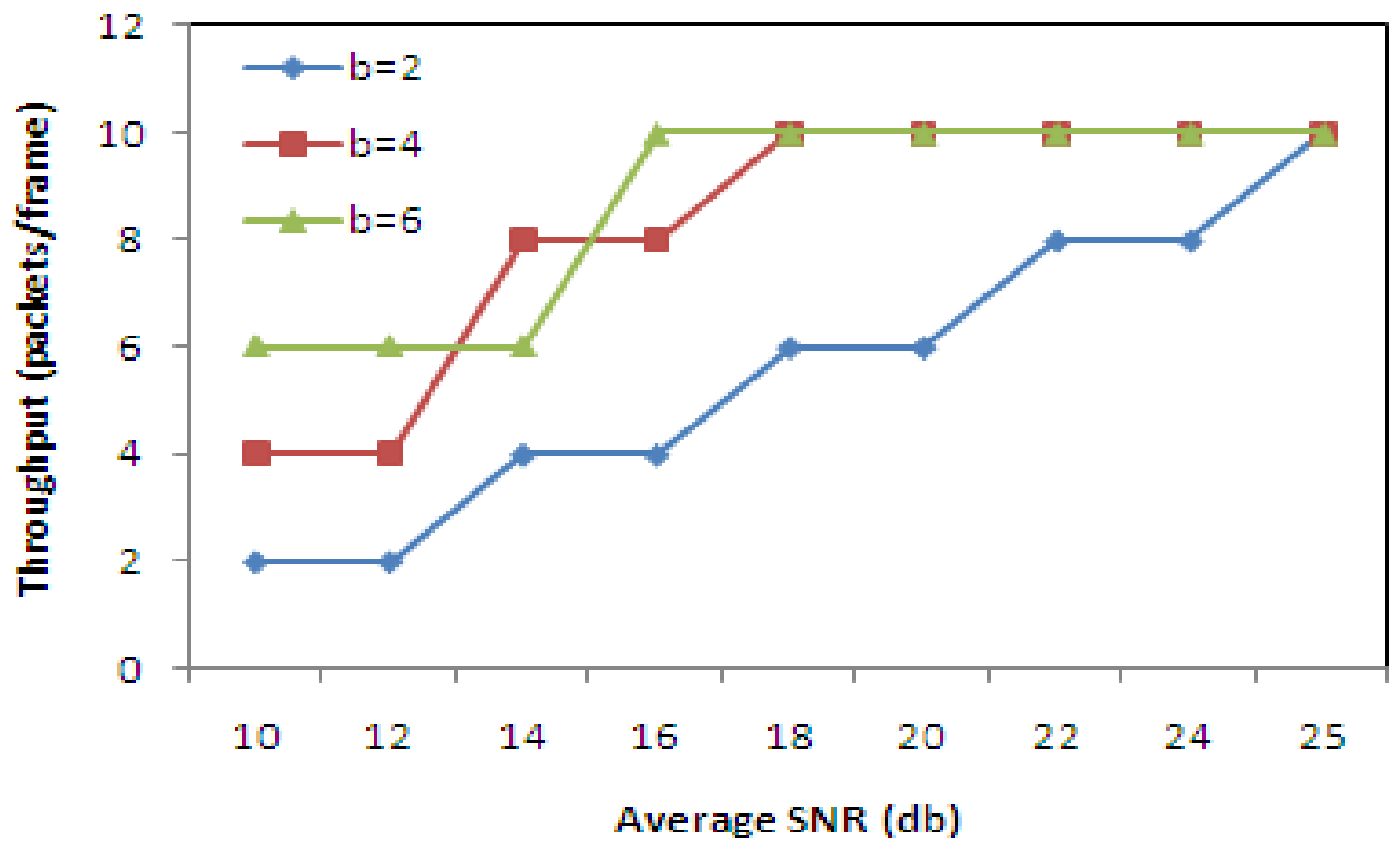

Figure 9: Throughput under different channel qualities.

When the channel quality improves, the transmitter can utilize higher modulation level and code rate to increase throughput (Figure 9). Note that if transmission rate is high enough to accommodate most of the arrival traffic, increased amount of allocated bandwidth or better channel quality will not impact the queue throughput since all the packets can be transmitted within a few frames. Moreover, different amount of allocated bandwidth results in different throughput.

## 8. Conclusion

In this paper, a queuing analytical model based on a Discrete-Time Markov Chain (DTMC) has been presented to analyze the packet-level performance in IEEE 802.16 broadband wireless access networks considering adaptive modulation and coding at the OFDMA physical layer. In the considered WiMAX system model, a base station serves multiple subscriber stations, and the base station allocates a certain number of sub-channels for each subscriber station.

To model the arrival process and the traffic sources we use the Batch Markov Arrival Process (BMAP), which enables more realistic and more accurate traffic modelling.

Using this queuing model, the impact of different traffic sources and the impact of channel quality on QoS parameters, such as average queue length, packet dropping probability, queue throughput and average packet delay, are analytically studied. Finally, the analytical results are validated numerically.

## Authors

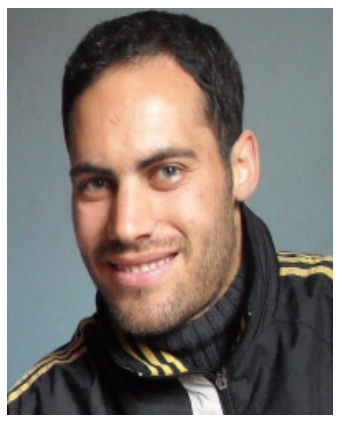

**Said EL KAFHALI** received the B.Sc. degree in Computer Sciences from the University of Sidi Mohamed Ben Abdellah, Faculty of Sciences Dhar El- Mahraz, Fez, Morocco, in 2005, and a M.Sc. degree in Mathematical and Computer engineering from Hassan 1st University, Faculty of Sciences and Techniques (FSTS), Settat, Morocco, in 2009. He has been working as professor of Computer Sciences in high school since 2006, Settat, Morocco. Currently, he is working toward his Ph.D. at FSTS. His current research interests performance evaluation, analysis and simulation of Quality of Service in mobile networks.

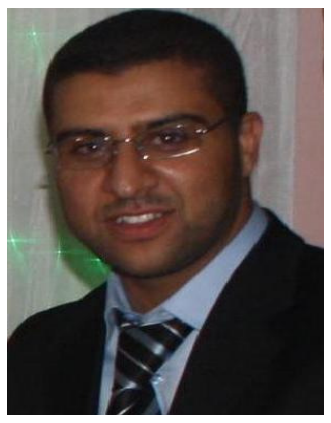

**Abdelali EL BOUCHTI** received the B.Sc. degree in Applied Mathematics from the University of Hassan 2nd, Faculty of Sciences Ain chock, Casablanca, Morocco, in 2007, and M.Sc. degree in Mathematical and Computer engineering from the Hassan 1st University, Faculty of Sciences and Techniques (FSTS), Settat, Morocco, in 2009. Currently, he is working toward his Ph.D. at FSTS. His current research interests include performance evaluation and control of telecommunication networks, stochastic control, networking games, reliability and performance assessment of computer and communication systems.

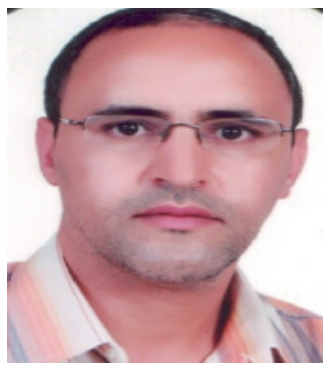

**Mohamed HANINI** is currently pursuing his PhD. Degree in the Department of Mathematics and Computer at Faculty of Sciences and Techniques (FSTS), Settat, Morocco. He is member of e-ngn research group. His main research areas are: Quality of Service in mobile networks, network performance evaluation.

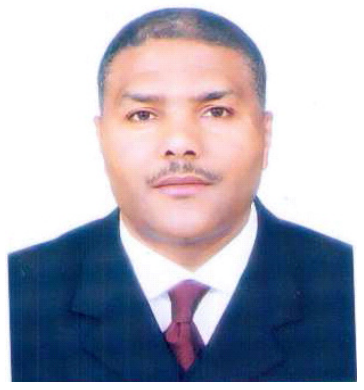

**Dr. Abdelkrim HAQIQ** has a High Study Degree (DES) and a PhD (Doctorat d'Etat) both in Applied Mathematics from the University of Mohamed V, Agdal, Faculty of Sciences, Rabat, Morocco. Since September 1995 he has been working as a Professor at the department of Mathematics and Computer at the faculty of Sciences and Techniques, Settat, Morocco. He is the director of Computer, Networks, Mobility and Modeling laboratory and a general secretary of e-NGN research group, Moroccan section. He was the chair of the second international conference on Next Generation Networks and Services, held in Marrakech, Morocco 8 - 10 July 2010. He is also a TPC member and a reviewer for many international conferences.

Professor Haqiq' interests lie in the area of applied stochastic processes, stochastic control, queueing theory and their application for modeling/simulation and performance analysis of computer communication networks.

From January 98 to December 98 he had a Post-Doctoral Research appointment at the department of systems and computers engineering at Carleton University in Canada. He also has held visiting positions at the High National School of Telecommunications of Paris, the universities of Dijon and Versailles St-Quentin-en-Yvelines in France, the University of Ottawa in Canada and the FUCAM in Belgium.